# Seamless Utilization of Heterogeneous XSede Resources to Accelerate Processing of a High Value Functional Neuroimaging Dataset

Don Krieger
*Neurological Surgery*
*University of Pittsburgh*
Pittsburgh, PA, USA
kriegerd@upmc.edu

Paul Shepard
*Physics and Astronomy*
*University of Pittsburgh*
Pittsburgh, PA, USA
shepard@pitt.edu

Ben Zusman
*Neurological Surgery*
*University of Pittsburgh*
Pittsburgh, PA, USA
zusmanbe@upmc.edu

Anirban Jana
*Pittsburgh Supercomputing Center*
Pittsburgh, Pa
anirbanjana@cmu.edu

David Okonkwo
*Neurological Surgery*
*University of Pittsburgh*
Pittsburgh, PA, USA
okonkwodo@upmc.edu

*Abstract*— We describe the technical effort used to process a voluminous high value human neuroimaging dataset on the Open Science Grid with opportunistic use of idle HPC resources to boost computing capacity more than 5-fold. With minimal software development effort and no discernable competitive interference with other HPC users, this effort delivered 15,000,000 core hours over 7 months.

The CamCAN Lifespan Neuroimaging Dataset [1,2], Cambridge (UK) Centre for Ageing and Neuroscience, was acquired and processed beginning in December 2016. The referee consensus solver deployed to the Open Science Grid [3,4] was used for this task. The dataset includes demographic and screening measures, a high-resolution MRI scan of the brain, and whole-head magnetoencephalographic (MEG) recordings during eyes closed rest (560 sec), a simple task (540 sec), and passive listening/viewing (140 sec). The data were collected from 619 neurologically normal individuals, ages 18-87. The processed results from both resting and task recordings are completed and available for download at http://stash.osgconnect.net/+krieger/ . These constitute ≈3.0 TBytes of data including the location within the brain (1 mm resolution), time stamp (1 msec resolution), and 80 msec time course for each of 3.7 billion validated neuroelectric events, i.e. *mean* 6.1 million events for each of the 619 participants.

Primary processing utilized the referee consensus solver deployed on the Open Science Grid (OSG). The pool of computing elements used by the solver was boosted using XSede allocations on Bridges (Pittsburgh Supercomputing Center) and Comet (San Diego Supercomputing Center). Computing elements were added directly to the OSG's HTcondor pool. This required no changes to the solver and only minor changes to the workflow management software which was fully operational in two days.

HTcondor's glidein mechanism is used to add the booster computing elements to the HTcondor pool. A script was deployed on each machine which queues Glidein's only when there are idle computing elements. Each Glidein's lifetime is limited to two hours, insuring that these jobs do not significantly delay other users of these important supercomputing resources. This opportunistic use of Bridges and Comet delivered more than a 3-fold increase in the number of core hours available for this effort.

*Keywords—magnetoencephalography, MEG, referee consensus, opportunistic computing, shared data, concussion, TBI*

## I. Introduction

It has been the informed expectation for a century that the keys to understanding the human brain will be found in measuring and understanding the electrical activity of neurons. Today, clinical neurophysiologists routinely measure single neurons to aide implantation of therapeutic devices deep in the brain [5]. Epileptologists use arrays of implanted "stereo EEG" electrodes and the population recordings obtained from them to diagnose and guide the treatment of intractable seizure disorders [6].

It is population activity which is thought to be the basis for brain function. Stereo EEG and comparable invasive methods produce voltage recordings with resolution of a few millimeters at best from up to a few hundred recording sites. Because the electric field interacts strongly with the conducting tissue in the brain, these measures are difficult to localize if there is much tissue between the field source and the electrodes. This problem is particularly pronounced when the recordings are made noninvasively from electrodes placed on the scalp.

Magnetoencephalography (MEG) provides an alternative noninvasive measurement approach with several advantages over scalp and even implanted EEG recordings. A typical MEG scanner is shown in Figure 1. The magnetic fields produced by minute electric currents within the brain are measured at the MEG sensor array with high fidelity. Unlike electric fields, magnetic fields do not interact with brain tissue; they pass through it as if it weren't there. In principle, the current which is the source of such a field is more readily localized. If the contribution to the MEG measurements of a single neuroelectric current source can be identified, the corresponding current can be accurately estimated using the Biot-Savart law [7].

The measurements at the MEG sensor array are due to an unknown and almost certainly large number of neuroelectric currents. The referee consensus method [8,9] enables reliable identification of one neuroelectric magnetic source at a time regardless of the number present. Since each identification is independent of all others, the method is readily ported to a loosely coupled supercomputing resource.

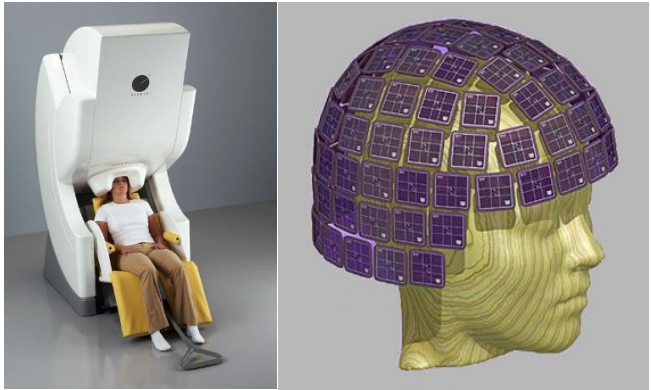

Fig 1. 306 channel whole head VectorView MEG scanner (Elekta, Inc., Stockholm, Sweden). The sensor array is composed of 102 planar "chips," each with 3 superconducting magnetic field sensors: one magnetometer, two figure-eight gradiometers at right angles to each other, and a superconducting quantum interference device (SQUID) coupled to each. The SQUID's are used to couple the minute currents induced by magnetic flux in the sensors to the room temperature electronics.

A solver based on this method has been deployed to the Open Science Grid (OSG), a large loosely coupled supercomputing resource which supports research in many domains including particle and nuclear physics, computational chemistry, genomics, proteomics, and neuroscience. The OSG is comprised of subsets of large computing systems primarily at colleges and national laboratories in the United States for periods when system managers anticipate low usage.

HTCondor[1] is used as the OSG's workflow management system; it receives job requests, distributes them to compute nodes, and returns results on job completion [3,4]. Transiently available compute nodes are added to the *condor pool* using HTCondor's lightweight glidein[2] mechanism. This adds the necessary configurations and daemons. A user with access to a set of compute nodes may run one or more glideins which add those nodes to the OSG condor pool and then run jobs on them by submitting the jobs to the OSG's instance of HTCondor.

Allocations have been provided through XSede[3] on Comet[4] and Bridges to add computing capacity for use on this project. HTCondor's glidein mechanism is used to incorporate this added capacity directly into the OSG pool as reserved job slots. This enabled adding transiently available XSede resources with almost no changes to the workflow management software developed within our group to handle data preparation, upload from within a hospital firewall, job spawning and monitoring, returned results post-processing, and download through the firewall. The shell scripts which manage the specialized portions of the workflow not readily handled by HTCondor are detailed below in **Workflow Management**.

II. METHODS: SOLVER IMPLEMENTATION AND RESERVED GRID COMPUTING RESOURCES

For this computing effort, the OSG condor pool is viewed as divided into two distinct pools. This required minor software changes. Jobs are designated for execution on one or the other pool when they are submitted to the OSG's HTCondor submit host. (1) The at-large pool is shared with all other OSG users. At-large OSG jobs run opportunistically per HTCondor's fair-share algorithm [10]. Our ongoing backlog of MEG recordings is queued to run at-large.

(2) The compute nodes attached via glideins running under our XSede allocations on Comet and Bridges are reserved. The CamCAN Lifespan Neuroimaging Dataset [1,2] is executed on these reserved nodes under a second OSG account number. This insures that the use of reserved compute nodes does not impact HTcondor's fair-share prioritization scheme as applied to at-large jobs. In addition, this enables the OSG to count the hours provided on the reserved nodes. These are not "billed" to our OSG allocation; instead they are billed to the Comet or Bridges allocations.

Over the 15 months that this has been operational, about 2/3 of the compute hours provided to our project has been via this reserve mechanism on Comet and Bridges with the bulk of it provided on Bridges, Figure 2.

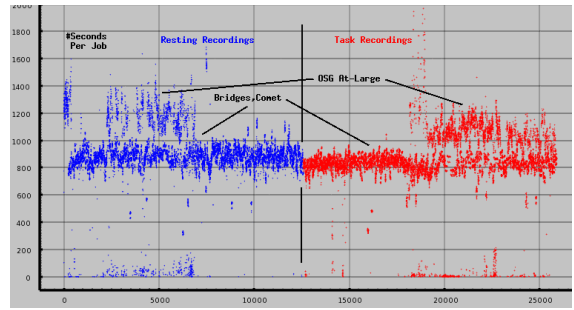

Fig 3. Jobs were run in groups of about 3000. Each dot in the figure represents completion of at least 99% of the jobs in a group. The x-axis spans 16 months. The y-axis indicates the average number of seconds required to complete a job for each group. Jobs run on Bridges or Comet required about 800 secs; jobs run on the OSG at-large required about 1050 sec. This was the time required to execute 619 subjects x 25 job groups per subject. That is about 15,000 job groups each for rest and task recordings with the total number of jobs at about 45,000,000 (see text).

Figure 3 shows that jobs run on the OSG at-large require about 30% more time to complete. Hence more than 70% of the processing of the CamCAN dataset was completed on these two machines. Without this resource, the processing would have taken four years instead of 16 months.

---

[1] HTCondor is the workflow management system which handles jobs submitted to the Open Science Grid. https://research.cs.wisc.edu/htcondor/
[2] Glidein is a mechanism by which one or more grid resources (remote machines) temporarily join a local HTCondor pool.
http://research.cs.wisc.edu/htcondor/manual/v7.6/5_4Glidein.html
[3] Extreme Science and Engineering Development Environment.
[4] Comet and Bridges are high performance computers situated at the San Diego and Pittsburgh Supercomputing Centers respectively.

To manage glidein spawning on Comet and Bridges, an independent software component was implemented in a script, *MonGlidein*. Slurm is used as the job management package on both systems which simplified the effort. Slurm commands were incorporated to poll the slurm queue and to spawn glideins.

The OSG as a whole utilizes computing resources which would otherwise be idle, *MonGlidein* embodies this same opportunistic principle. (a) Glideins are queued on Bridges with "deferred" priority, i.e. each is assigned a compute node only if no jobs are pending in the slurm queue. The Comet system administrators have chosen to not implement this measure.

(b) Glideins run for two hours at a time. This insures that no job queued by another user will wait on completion of a glidein for long. The dead time at the end of the short glidein lifespan during which new OSG job starts are prohibited reduces job slot occupancy to 94%. Glideins with a 3-hour lifespan show 98% occupancy. This small decrement in efficiency was accepted to improve the responsiveness of the system. Furthermore, this short run time promotes effective utilization of backfill cycles, i.e. cycles on large groups of cores which must drain for hours to accumulate the resources requested by large multi-node jobs.

(c) On Comet, only one glidein is allowed to remain in the

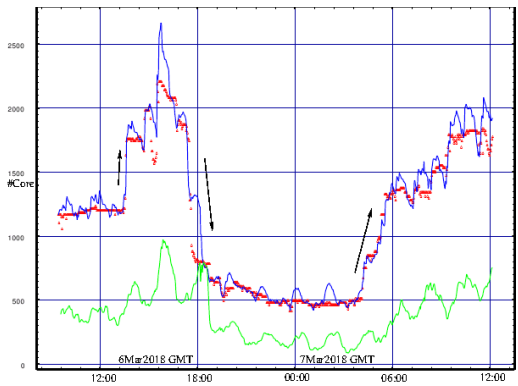

Fig 4. Opportunistic spawning and execution is responsive (a) to demands by other users (down-arrow) and (b) to the availability of idle cores (up-arrows). Bridges reserved jobs slots (red) and occupied jobs slots (blue) are shown for a 27-hour period. See text. Live versions of this figure may be found **online**.

slurm queue at a time. This insures that when groups of compute nodes go idle, a glidein will compete with other jobs for only one. This is measure is redundant on Bridges due to measure (a), deferred priority.

The responsiveness of the system is shown in Figure 4. The 2-hour glidein lifespan produces the sharp response to jobs spawned by other users, down arrow in the figure. The response to compute node availability is also sharp, up arrows in the figure.

*MonGlidein* polls the slurm queue and spawns a group of three glideins only if no glideins are found waiting. When a new group is spawned, *MonGlidein,* pauses for 30 seconds and checks to see if the group has begun running. If it has, a new group is spawned. If not, the wait time is multiplied by 1.5. The wait time for polling is limited to 300 seconds.

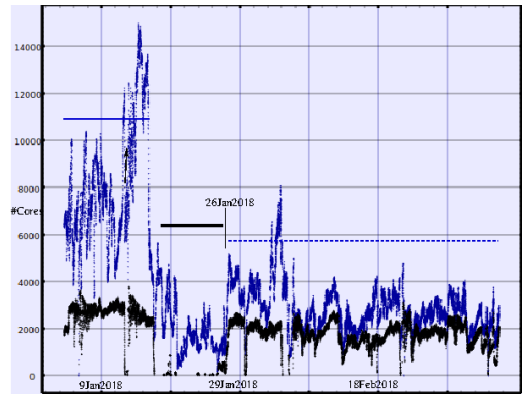

Fig 2. OSG core usage (navy) and Bridges core usage (black) are shown for 60 days. Comet was not available during this period. Bridges usage comprised about 43% of the total. The at-large OSG usage is the difference between the traces. At-large usage was high for the first 10 days (blue bar) due to the typical New Year's lull in use by other groups. A major Bridges maintenance period followed (black bar). The final 40 days (dotted bar) demonstrate a more typical Bridges/at-large usage, i.e. 62.6%. This percentage was 61.6% for the 448 days from the first day of Bridges use, 13Dec2016, to 7Mar2018. See text.

Each glidein utilizes a fraction of the cores on a compute node. For "regular memory" nodes, 12 of 24 cores are used on Comet, 14 of 28 on Bridges. Node "sharing" avoids reduction in priority by slurm for requests for complete compute nodes. It also is much more effective in utilizing idle cores during periods of high utilization by other users. Figure 5 shows a rapid burn through 5000 core hours on Bridges regular memory nodes during such a period.

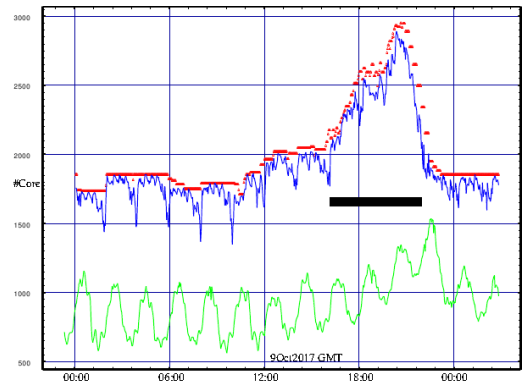

Fig 5. Bridges reserved jobs slots (red) and occupied jobs slots (blue) are shown for a 30-hour period. The black bar indicates a period during which 5000 "backfill" core-hours were used on regular memory nodes. Large memory nodes were also utilized throughout this period. Live versions of this figure may be found **online**.

The calculations are performed by a compiled and static linked executable. A single instance requires ≈300 Mbytes of memory and ≈35 Mbytes of disk space and network I/O. This network I/O is potentially a problem since the number of solver instances, i.e. jobs, required to process a single 560 second recording block is ≈75,000. This could require up to 2.6 TBytes of network I/O.

Two mechanisms are used to reduce this load. (A) For jobs run at-large on the OSG, data transfers are mediated by http servers which incorporate the SQUID data caching proxy. This is effective because the data is identical for all jobs processing identical time segments, i.e. groups of ≈3000 jobs. Numerous spot checks show a consistent "hit" rate well above 95% which reduces the required data movement by more than 20:1. (B) For jobs which run on reserved glideins on Comet or Bridges, a single copy of the data is placed on the scratch disk of the machine for each group of ≈3000 jobs.

CreateForward is a python script which generates the forward solution matrix, i.e. the map:

$$\text{neuroelectric current} \rightarrow \text{magnetic field}.$$

A single CreateForward instance with accompanying python package requires ≈400 Mbytes of memory and 20 Mbytes of disk space and network I/O. CreateForward is identical for all jobs. The SQUID file cache hit rate is near 100%. For reserved glideins on Bridges or Comet, this package is handled in the same way as is the data.

### III. METHODS: WORKFLOW MANAGEMENT

Shell scripts were developed to manage the workflow. This scripting effort was and is guided by several principles: (1) The workflow is conceived as a real-time problem. (2) The software pieces are each viewed as evolving prototypes [11]. In almost all cases, modifications have been introduced "on the fly" without disrupting the workflow. (3) Each script must be as lightweight as possible. Attention is particularly paid to minimizing use of shared resources, e.g. network and disk I/O. (4) As faults occur, the error conditions are identified and analyzed. The scripts are then modified to avoid or trap those error conditions, i.e. the software is progressively "bullet-proofed."

As is usually the case with supercomputing resources, access to the OSG is via a front-end node. In our case, this is a machine at Indiana University provided by XSede, *xd-login*. The solver processing pipeline begins with preprocessing the raw MEG files and uploading them to *xd-login;* it ends with the return of results files from *xd-login*. Both of these data movement operations are handled from the *master* node which is behind a firewall with no firewall exceptions required.

The parallelism inherent in the referee consensus method is due to the fact that one spatial location is tested for the presence of a neuroelectric current for one data segment at a time. Each search is independent of all others which freely enables separating searches both by time segment and by spatial location. In practice, one search is conducted per data segment for each 8x8x8 mm brain voxel. Fully covering the volume of the brain, ≈1.5 liter, requires ≈3,000 voxels.

Each CamCAN resting recording was collected in a single continuous 560 second sitting. This is divided into 25 time blocks; each solver instance searches one voxel for one of these time blocks. Hence ≈75,000 solver instances, i.e. jobs, are required to process one 560 second resting recording. These parameters were selected to limit the run length for each job to a *mean* of 850 sec. This *mean* runtime is long enough to run efficiently on the OSG and short enough to minimize the lost cycles due to idle job slots at the end of a 2-hour glidein run.

Each solver instance is spawned to run on a separate processor/core, i.e. each solver instance is a separate OSG job. These jobs are assembled and spawned by a script which runs on *xd-login*. A single *spawner* instance handles these functions for each group of five time segments for the ≈3000 voxels covering the whole brain. Hence five *spawners* are used for each 560 second resting recording, each of which handles ≈15,000 jobs. At the end of its run, each *spawner* spawns a *collector* script. The *collector* collects and sorts the results files and then sets a flag which is detected by the script on the *master* node which fetches the results.

There is a single *submaster* script running on *xd-login* which spawns the *spawners.* The *submaster* polls the HTcondor queueing system and provides job group completion information required by all *spawners* and *collectors.* Polling HTcondor for this information, a relatively heavyweight operation, is thereby confined to a single process which does so every 3 minutes. The *submaster* also controls the number of data blocks which have been uploaded to *xd-login* by the *master* node. It does so by setting a GO/STOP flag which is detected by the data preprocessing script which is running on the *master.*

This workflow management and spawning system works without interfering with other workflows with continuous *xd-login* throughput as high as 35,000 jobs for extended time periods. The system typically runs without fault for a month or more, i.e. from one *xd-login* reboot to the next.

### IV. METHODS: HOW THE SOLVER WORKS

The magnetic field strength at each of the 306 MEG sensors is the weighted sum of an unknown but likely large number of magnetic fields. A few of these such as mains powerline signals may be identified and subtracted off, filtered out, or otherwise eliminated by standard signal processing methods. What remains is an unknown number of magnetic field signals, many of which are of significant interest. The challenge is to identify as many of these as possible, and to identify both the location and time course of the electric current which is their source. Hence, we have a deconvolution problem for which both the number, location, and time course of the contributors to our measurements are unknown.

Note that the nature of the problem as one requiring deconvolution may be ignored for data in which the magnetic field strength from a single current constitutes the great majority of the total measured field, e.g. very high amplitude epileptiform activity. This scenario can be forced by reducing the data to averaged MEG signals for which the time segments contributing to the average are time locked to a repeated event, e.g. a button press or stimulus presentation. This enhances the signal/noise of excursions in the waveforms which are time-locked to the sync point and so reduces the sensitivity of the processing method to non-sync'd signals. It also markedly attenuates high frequency activity and averaging "collapses" the data by the number of time segments that are averaged. These consequences of averaging typically reduce the quantity of information that can be extracted from the record by 100:1 or more.

In general, because the number of contributors to the measured field is unknown, it is not possible to solve the deconvolution problem directly, i.e. to find and characterize the complete set of electric currents whose magnetic fields account for the measurements. However one can write a set of equations which contain terms for a large number of putative neuroelectric currents whose locations cover the volume of the brain as densely as wished. For this approach, the number of unknown parameters is far larger than the number of measurements, i.e. the solution is "ill-posed." A unique and meaningful solution may still be found by introducing a regularization term and solving using a pseudo-inverse, e.g. MNE [12], sLoretta [13]. These approaches provide limited information as (1) they are susceptible to interference and (2) they have low spatial resolution.

An alternative approach is to construct a digital filter for each location within the brain whose activity is of interest, e.g. SAM [14]. This approach produces an estimate for the electric current at a single location but that estimate is sensitive to sources found elsewhere which "leak through" the filter and contaminate it. This compromises both the spatial resolution and fidelity of the amplitude estimate. In addition, neither this method or the others provide a reliable and robust validation scheme.

The referee consensus method uses a large family of digital filters to test and validate the presence of one neuroelectric current at a time. A cost function and associated probability are computed which are robust enough to use $10^{-12}$ as the threshold p-value to accept a current at a specific location as a true source of a detectable magnetic field. These capabilities are used to conduct a search of the brain volume for one 80-msec data segment at a time as detailed below and in the appendix. The effectiveness of this scheme in rejecting false signals during noisy data epochs is shown in Figure 6.

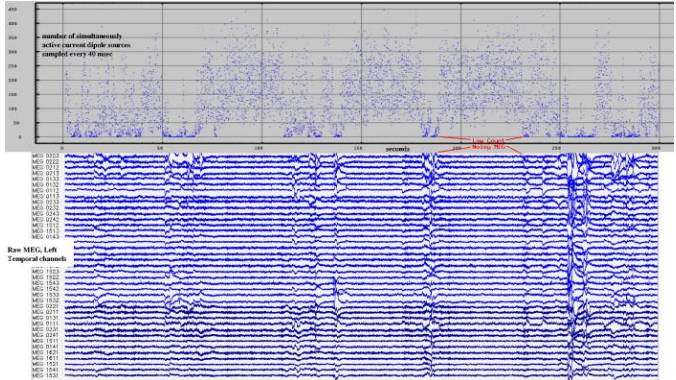

Fig 6: Neuroelectric source validation at $p < 10^{-12}$. The referee consensus solver automatically fails when the recordings are noisy. 300 seconds of raw MEG (lower) and neuroelectric currents (upper) are shown. The number of validated ($p < 10^{-12}$) currents identified drops markedly when the MEG is noisy.

Each data segment is a sequence of 80 observations in each of 306 magnetic field sensors. This may be considered as a vector with 80 x 306 = 24,048 components, i.e. as a point in a real space with 24,048 dimensions. The contribution to the magnetic field measurements at 306 sensors due to a dipole electric current at location **X** is the resultant magnetic field as it waxes and wanes over the 80 msec observation period.

The forward solution [15] which defines this relationship is nonlinear in the location parameters, xyz, but is linear in the amplitude. Hence the measured field due to the current at **X** is readily conceived as a point in $\mathcal{R}^{306} \times \mathcal{R}^{80}$, i.e. the Cartesian product of two real spaces with 306 (field shape) and 80 (amplitude time course) dimensions.

For a fixed **X**, the "shape" of the field is fixed, i.e. the ratios between each component of the field and every other is constant. Therefore the 80-point time course of the field measurement at each sensor is perfectly correlated or anti-correlated with that at every other sensor with the exception of sensors whose sensitivity is zero. That shape produces a set of 306 <u>known</u> values at the 306 magnetic field sensors, each of which is multiplied by an amplitude at each time point. Because the shape of the field is fixed, that shape multiplied by the amplitude always falls within a 1-dimensional subspace of $\mathcal{R}^{306}$, i.e. on a line within the space.

What is <u>unknown</u> is the 80-point time course of the current amplitude and the resultant 80-point time course of the magnetic field amplitude. It is the amplitude at each time point which determines where on the line in $\mathcal{R}^{306}$ the shape falls.

The 2-fold task of the solver is (1) provide a robust measure of confidence that a dipole current is detected at location **X** and (2) estimate the time course of the current amplitude. The method is briefly presented in the Appendix. It uses a family of digital filters which are highly tuned to the 1-dimensional subspace of $\mathcal{R}^{306}$ where the field due to a current at **X** falls and detuned to the orthogonal $\mathcal{R}^{305}$ subspace. Using this family of specialized filters produces a robust probabilistic measure of confidence that current is or is not present at **X** and a high fidelity estimate of the time course of the current when present.

## V. RESULTS

The referee consensus solver applied to MEG recordings isolates and identifies one localized neuroelectric current at a time, regardless of how many currents are simultaneously present. The solver is applied to the raw data stream one 80 millisecond (msec) time epoch at a time. For each time epoch, the total brain volume is searched for neuroelectric currents. In the control cohort resting data, a *mean* of 440, *sd* 130, simultaneously active currents were identified for each time epoch. The search progresses through the data stream in 40 msec steps, i.e. 25 steps per second, yielding about 11,000 neuroelectric currents per second, each validated at $p < 10^{-12}$. This threshold insures that almost all of the identified currents are real.

Here are several key characteristics of the neuroelectric current measures extracted from the MEG using the referee consensus solver. Note that the MEG is sampled at 1 KHz and filtered with high and low pass at 10 and 250 Hz, 5 Hz roll-off. Mains noise is removed at 50, 100, 150, 200, and 250 Hz. Continuous head positioning is recorded and used to correct the spatial localization once per second. Each identified neuroelectric current is an 80-point (80 msec) waveform localized with 1 mm resolution and validated at $p < 10^{-12}$. The frequency content of the identified waveforms is dominated by activity 20-150 Hz.

Note that those aged 18-65 were used to represent the portion of the CamCAN which matches our other cohorts. Hence the *n* is 411 rather than 619.

Approximately equal numbers of neuroelectric currents are identified in both gray and white matter. The unexpected detection of neuroelectric currents in the white matter and the high spatial resolution of their localization enables robust measures of gray vs white differential activation. Figure 7 shows a typical example from an individual in which differential activity, gray vs white, is seen with $p < 10^{-16}$ in several regions. Note that the gray matter region is more active than the white as expected in the right inferior parietal pair but the differential is reversed for the right rostral middle frontal regions.

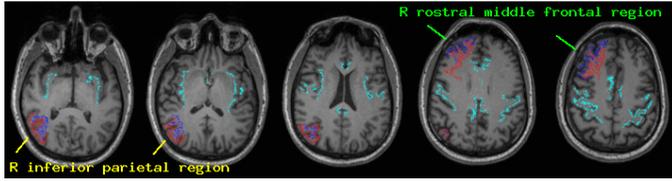

Fig 7. Activation is compared between the right inferior parietal cortex and the adjacent white matter (yellow marks) and between the right rostral middle frontal cortex and the adjacent white matter (green marks). The $\chi^2$ statistic is computed for the difference between the densities. The fully saturated color indicates that the comparison is significant at $p < 10^{-16}$. For the parietal region the ROI shown in red, the cortex, is as expected, more active than the adjacent white matter, shown in blue. For the frontal region, the differential activation is the opposite, i.e. the adjacent white matter is more active than the cortex.

The 560 second CamCAN resting recordings produced a *mean* of 6.1 million (*sd* 1.9 million) localized and validated current estimates for each subject. These profuse, high resolution, highly reliable measures present a strikingly detailed and dynamic characterization of brain activity. The full set of results for the 619 members of this neurologically normal cohort represents an invaluable scientific resource. It contains measurements from 619 men and women, ages 18-87, providing age and sex matches for any adult population. The results have been placed online including the following:

(1) All validated neuroelectric sources for each member of the CamCAN cohort. Each source listing includes the xyz coordinates (1 mm resolution) of the source location, a time stamp (1 msec resolution), and the 80-point time course of the current amplitude.
(2) The output of the freesurfer run for each member of the cohort.
(3) The list of coordinates on a 1 mm grid which were searched for currents and the freesurfer region in which each location resides.
(4) A tool which is usable through the web browser to generate a complete list of download files for each member of the cohort. That list is designed to be used with wget to download the data.
(5) A tool usable on a linux machine with Intel or AMD processor which assembles the list of currents into simultaneously active groups and writes them to standard output as ascii characters. This need not be used since the data is already in compressed ascii form. But for many operations it will prove convenient.

Rather than pursuing analyses which capitalize on the high time resolution of our measures, our emphasis has been on the reliability of the measures. We use counting, i.e. the number of identified currents per unit volume as a measure of activity. We use freesurfer [16] to identify 164 standard brain regions of interest (ROI) and compute an activity density measure for each. The density measure is normalized both by the total counts[5] and total brain volume:

$\rho_{ROI} = (count_{ROI}/count_{total})/(vol_{ROI}/vol_{total})$

The counts are so large and the false identification rate is so low (e.g. Figure 6) that we have ample statistical power using the $\chi^2$ statistic to identify differences between regional density measures, e.g. cortical region vs adjacent white matter region, region during rest vs the same region during task. The activity densities are reduced in deep structures and in the cerebellum. This is likely due to reduced sensitivity because these structures are further from the MEG sensor array. However, the numbers are ample even in the brainstem to see significant differential activation ($p < 10^{-4}$) when comparing rest and task. Global patterns identified by standard eigenvector methods enable discrimination between the CamCAN cohort and those with chronic symptoms of concussion or with those at-risk for HIV with 100% accuracy (Figure 8). And the patterns themselves, also shown in the figure, are of significant interest.

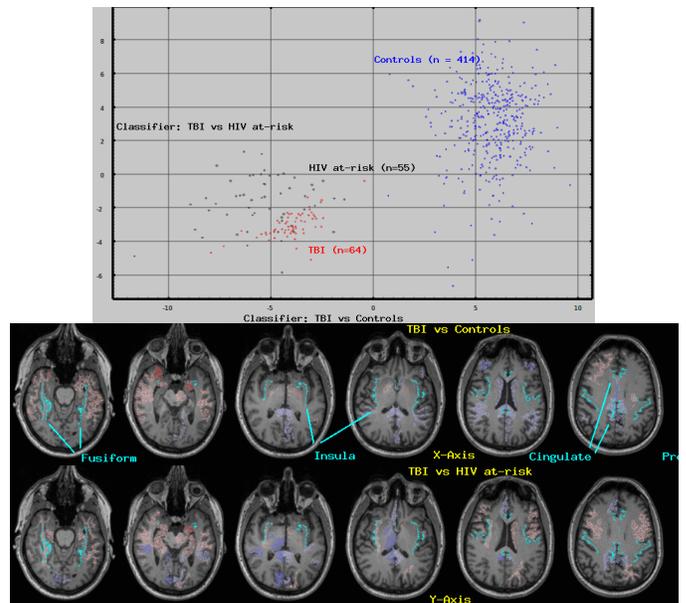

Fig 8: Regional patterns of brain activity[6] demonstrate high classification accuracy between cohorts. Neuroelectric activity density ($\rho_{act}$) was measured from resting MEG recordings in each of 164 brain regions in 3 experimental cohorts, (1) controls (n=414), (2) TBI (n=64), (3) HIV at-risk (n=55). The middle and lower panels show the brain regions which contribute their fair share or more to the classifier; Red/blue indicates a positive/negative weight, the more saturated (deeper) the color, the greater the weight. The cyan landmarks are the boundaries between gray and white matter in the pre-central, cingulate, insula, and fusiform regions.

We have also demonstrated a regional measure which provides network connectivity information. Instead of counting the number of neuroelectric currents which occur within a region, we count the number of pairs of simultaneously occurring currents for which one of the pair occurred within the region. Since we know the incidence over the recording session of each member of the pair, we can compute the chance that we would count the number of coincidences. We set a threshold of p < 10$^{-8}$ to accept a pair as occurring more often than expected. For the control cohort, we count a *mean* of 114,399 pairs for each 560 second resting recording, *sd* 80,492, *min* 8,328, *max* 563,220.

---

[5] Dividing the counts for a region by the total count normalizes the numerator of the "density" for variations due to regional volume, data quality and record length
[6] The first 20 independent components (ICAs) of these measurement vectors were identified from the control cohort. Each person's 164 measures were reduced to their factor scores on these 20 ICAs. Stepwise linear discriminant analysis (BMDP7M) was used on the factor scores to identify the best linear classifiers for one cohort from another. The classification accuracy for TBI vs controls was 100%; accuracy for TBI vs HIV at-risk was 76%.

The number of such pairs with one member of the pair within a region and the other member outside the region provides a measure of the "coupling strength" of the region to the rest of the brain. The regional density is computed from this count in the same way as the activity density described above and, in fact, can be compared directly with it, i.e. the activity density of a region can be compared with its coupling strength.

Both the activity density and coupling strength measures provide enormous statistical power to test hypotheses about select brain regions in a single individual as shown in Figure 9. The figure shows differences between cortex and adjacent white matter for both activity and coupling measures for the entire CamCAN cohort. In addition, since the solver enables formal comparison of the activity densities of a cortical region with the adjacent white matter region. Under the plausible assumption that the rim of white matter largely terminates in the adjacent gray matter, the ratio of these densities, $\rho_{ctx}/\rho_{wm}$, is a measure of cortical excitability. This therefore provides a promising and previously inaccessible tool in addition to activity and coupling strength.

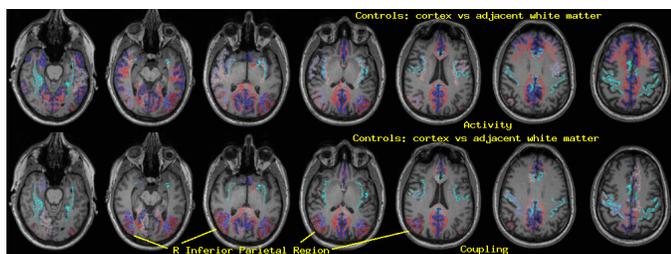

Fig 9: Regional differential activation/coupling, control cohort, cortex vs adjacent white matter. The average activity and coupling densities (see text) were computed for each of 164 regions for the control cohort (n=414). The mean for each cortical region was compared with the mean for the adjacent white matter region using Welch's t-statistic. Red/blue regions are more/less active (upper panel) or coupled (lower panel) than the adjacent region. The p-value for each regional comparison ranges from p < 10$^{-4}$ (light color) to p < 10$^{-8}$ (fully saturated color). Most of the regions which show differential activation show greater activation in the white matter than in the adjacent cortex. The right inferior parietal region is highlighted because its differential coupling strength is much more significant than its differential activation, despite the reduced statistical power in the coupling measure.

## VI. DISCUSSION

This work is based on resource sharing and on the complementary principle, opportunism. The key elements of the work are shared, i.e. the CamCAN lifespan normative dataset, the Open Science Grid, the high performance computing resources, and the results of the work. Absent the data or the supercomputing resources to process it, the work could not go forward. Absent sharing the results, the value of the work is stunted since these results provide a characterization of the brain activity of a neurologically normal population with unprecedented detail. The availability of this dataset and the processed results provides controls to the wide community of workers using MEG to study both normal and pathological human brain function.

The emphasis of the work is methodological. The challenge is to use the measurements to advance our understanding of brain function both for research and for clinical purposes. The information extracted by the referee consensus solver provides unique opportunities to accomplish this.

(1) The individual neuroelectric currents are localized with millimeter precision. Their time course is measured with high fidelity in the frequency range: 12 – 250 Hz.

(2) Analysis of regional current density provides extraordinary statistical power in measuring differential regional activation, e.g. Figures 7 and 9.

(3) This approach demonstrates the capability of the solver to reliably identify neuroelectric currents within the brain's white matter, e.g. Figure 7. Aside from a few recent results in the fMRI literature [17], this is the only known method which produces detailed functional measures from this critically important and previously inaccessible portion of the brain.

(4) The robust detection of differential activation between a cortical region and the adjacent white matter rim and the fact that it is often the white matter which shows greater current density suggests that it is synchronous volleys of action potentials which are at the basis of the currents detected by the solver. This conclusion is supported by the predominance of high frequency content in the identified waveforms.

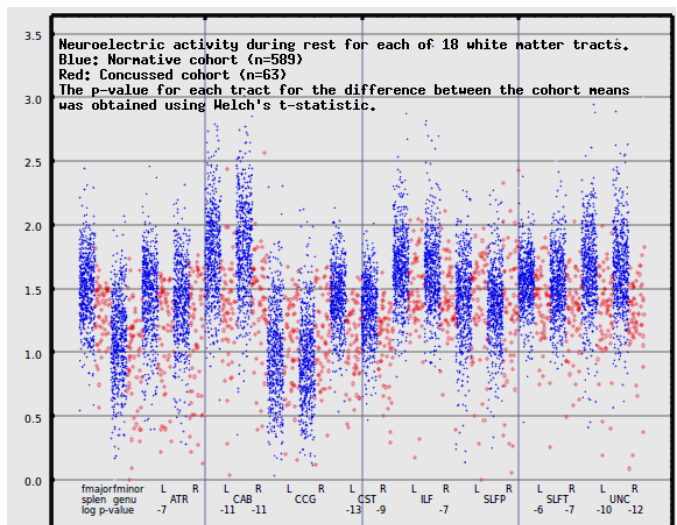

Fig 10. Functional activity in 18 automatically identified tracts is shown in blue for each member of the neurologically normal cohort, ages 18-88, n = 589, and in red for each member of the concussed cohort, n = 63.

The measures were obtained from resting MEG recordings (see text). The xyz coordinates of the volume occupied by each tract were estimated using tracula (Martinos Brain Imaging Center, Harvard University). Neuroelectric activity density was computed and is interpretable as a measure of connectivity strength across the tract.

The p-values near the bottom of the figure indicating significant differences (p < 10-6) for the indicated tracts. Note that for each significant finding, the mean activity for the concussed cohort is less than that for the normative cohort.

(5) The adjacent rims of white matter paired with cortical regions by freesurfer are at most 5 mm thick. They include the thin layer of white matter extending into the cortical gyri and separating the cortex on one side of the gyrus from the other. Under the assumption that the majority of the fibers in each white matter region terminate in the adjacent cortex, we propose the ratio of the activity in a cortical region with the adjacent white matter as a measure of cortical excitability.

(6) Analysis of regional activity using standard dimensional reduction methods, i.e. eigenvector analysis, produces physiologically interpretable patterns which distinguish between groups with very high accuracy, e.g. Figure 8, and which are highly correlated with a variety of pathological symptoms. The identified patterns provide one dimensional axes within the very high dimensional space of brain states with a defined direction toward "normal."

(7) Analysis of groups of simultaneously active currents provides remarkable statistical power to detail the coupling strength, i.e. functional connectivity, between one brain volume and the rest of the brain, e.g. Figure 9.

(8) Analysis of the regional current density over time provides a means to extend the utility of these processed results to study low frequency phenomena. The presumed interpretation of the extracted current waveforms as representative of synchronous volleys of action potentials ties this analysis directly to the underlying neurophysiology.

Transcranial magnetic stimulation (TMS) [18] applied to select brain regions is a promising treatment modality. This safe and painless device has been used successfully in patients with major depression and obsessive compulsive disorder. The activity and coupling measures described here will likely be sensitive to treatment with TMS. Hence these measures might be used to monitor changes in brain activity in those undergoing this treatment. Since measures on our large control cohort nominally provide the direction of asymptomatic, if we can understand and predict how TMS effects brain activity, these measures could potentially be used to guide treatment with TMS.

ACKNOWLEDGMENTS

We gratefully acknowledge the invaluable contributions made to this effort by the Department of Defense, the Cambridge (UK) Centre for Ageing and Neuroscience, the Extreme Science and Engineering Development Environment (Xsede), the Open Science Grid (OSG), the San Diego Supercomputing Center, the Pittsburgh Supercomputing Center, the XSede Neuroscience Gateway, Mats Rynge, Rob Gardner, Frank Wurthwein, Derek Simmel, and Mahidhar Tatineni. Data used in the preparation of this work were obtained from the CamCAN repository [1,2]. The OSG [3,4] is suppored by the National Science Foundation, 1148698, and the US Department of Energy's Office of Science.

SUPPORTING INFORMATION

http://stash.osgconnect.net/+krieger/S1_CamCAN_Atlas.zip
This archive contains several of cohort-wide atlases and sample summary results from one individual. It also contains a typical output file created by the referee consensus solver. This file contains the particulars from the 13,758 80-msec neuroelectric current waveforms identified in one second from this same individual's resting MEG recordings.

REFERENCES

[1] JR Taylor, NWilliams, R Cusack, T Auer, MA Shafto, M Dixon, LK Tyler, Cam-CAN, RN Henson, "The Cambridge Centre for Ageing and Neuroscience (Cam-CAN) data repository: Structural and functional MRI, MEG, and cognitive data from a corss-sectional adult lifespan sample," Neuroimage 144: 262-269, 2015, PDF.

[2] MA Shafto, LK Tyler, M Dixon, JR Taylor, JB Rowe, R Cusack, AJ Calder, WD Marslen-Wilson, J Duncan, T Dalgleish, RN Henson, C Brayne, CamCAN, FE Matthews, "The Cambridge Centre for Ageing and Neuroscience (CamCAN) study protocol: a cross-sectional, lifespan, multidisciplinary examination of healthy cognitive ageing. BMC Neurology 14(204), 2014, PDF.

[3] R Pordes, D Petravick, B Kramer, D Olson, M Livny, A Roy, P Avery, K Blackburn, T Wenaus, F Wurthwein, "The open science grid," J Physics 78, 2008, PDF.

[4] I Sfiligoi, DC Bradley, B Holzman, P Mhashilkar, S Padhi, F. Wurthwein, "The pilot way to grid resources using glideinWMS," WIR World Cong Comp Sci and Info Engg 2: 428-432, 2009, PDF.

[5] I Telkes, J Jimenez-Shahed, A Viswanathan, A Abosch, N Ince, "Prediction of STN-DBS electrode implantation track in Parkinson's Disease by using local field potentials," Front. Neurosci 10: 1-16, 2016, **PDF**.

[6] F Moroni, L Nobili, G Curcio, F De Carli, F Fratello, C Marzano, L De Gennaro, F Ferrillo, M Cossu, S Francione, G Lo Russo, M Bertini, M Ferrara, "Sleep in the human hippocampus: a stereo-EEG study," PLoS One 2(9): e867, 12Sep2007, PDF.

[7] J Sarvas, "Basic mathematical and electromagnetic concepts of the biomagnetic inverse problem," Phys. Med. Biol. 32(1): 11-22, 1987, PDF.

[8] D Krieger, M McNeil, J Zhang, W. Schneider, X Li, DO Okonkwo, "Referee consensus: a platform technology for nonlinear optimization," XSEDE 2013, ACM 978-1-4503-2170-9/13/07, PDF.

[9] D Krieger, M McNeil, J Zhang, A Puccio, W Schneider, X Li, DO Okonwko, "Very high resolution neuroelectric brain imaging by referee consensus processing," Intl. J. Advd. Comp. Sci. (4)1: 15-25. Jan 2014, PDF.

[10] B Jayatilaka, T Levshina, M Rynge, C Sehgal, M Slyz, "The OSG open facility: A sharing ecosystem," J Phys: Conf Ser. 664 PDF .

[11] M Deutsch, "Software Quality Engineering: A Total Technical and Management Approach," Prentice-Hall, Englewood Cliffs, NJ, 1988.

[12] MS Hamalainen, RJ Ilmoniemi, "Interpreting magnetic fields of the brain: minimum norm estimates," Med. and Biol. Engg and Compg 32(1): 35-42, 1994, PDF.

[13] RD Pascual Marqui, CM Michel, D Lehmann, "Low-resolution electromagnetic tomography – a new method for localizing electrical activity in the brain," Intl J Psychophys 18(1): 49-65, 1994, PDF.

[14] J Vrba, SE Robinson, "Signal processing in magnetoencephalography," Methods 25: 249-271, 2001, PDF.

[15] J Sarvas, "Basic mathematical and electromagnetic concepts of the biomagnetic inverse problem," Phys. Med. Biol. 32(1): 11-22, 1987, PDF.

[16] B Fischl, M Sereno, A Dale, "Cortical surface-based analysis," NeuroImage 9: 195-207, 1999, PDF.

[17] T Warbrick, J Rosenberg, NJ Shah, "The relationship between BOLD fMRI response and the underlying white matter as measured by fractional anisotropy (FA): a systematic review," NeuroImage 153: 369-381, 2017, PDF.

[18] M Eid, Y Qanir, "Evidence based practice: transcranial magnetic stimulation for major depression," Intl. J. Psych. Br. Sci. 1(1): 13-20, 2016, **PDF**